\documentstyle[aps,multicol,epsf]{revtex}

\begin{document}
\draft
\title{Short time dynamics of a two-dimensional majority vote model}

\author{J. F. F. Mendes and M. A. Santos}

\address{
Departamento de F\'\i sica and Centro de F\'\i sica do Porto, Faculdade de Ci\^encias, Universidade do Porto\\
Rua do Campo Alegre 687, 4150 Porto -- Portugal
}

\maketitle

\begin{abstract}
Short time Monte Carlo methods are used to study the nonequilibrium ferromagnetic phase transition in a majority vote model in two dimensions. The existance of an initial critical slip regime is verified. The measured values of dynamic exponents $z=2.170(5)$ and $\theta = 0.191(2)$  are in excellent agreement with those of the kinetic Ising model universality class.
\end{abstract}

\pacs{PACS numbers: 05.70.Ln, 05.50.+q, 64.60.Ht, 02.70.Lq}

\begin{multicols}{2}

\narrowtext

The identification and caracterization of universality classes for nonequilibrium systems are far less settled than in the case of systems in thermal equilibrium. Nevertheless, many model systems with microscopic irreversible dynamic rules (no detailed balance) and two states per site have been found to fall in the Ising model universality class, as far as the static behaviour is concerned [1-8]. Grinstein, Jayaprakash and Yu He \cite{GJY} have argued that, provided the rules are up-down symmetric, both the statics and the (long time) dynamics of those models are the same as those of the kinetic Ising model.
We are however not aware of any direct determination of the dynamic $z$-exponent for such models. In the present work we have applied the recently proposed early time dynamic Monte Carlo technique  \cite{janssen89}, \cite{Li95}, \cite{Li96} to investigate the dynamic behaviour of a model in the above mentioned class - the majority vote model (whose rates can be seen as a combination of two Glauber dynamics in contact with two heat baths at different temperatures) \cite{liggett}, \cite{mario}. Previous studies of short time dynamics were concerned either with equilibrium systems, like Ising \cite{Li94}, \cite{Li95}, \cite{Li96} or Potts \cite{schulke}, \cite{okano97}, \cite{z97} models, or with a nonequilibrium phase transition in a distinct universality class \cite{nora96}.

Janssen, Schaub and Schmittmann \cite{janssen89} have shown that, when a system with relaxational dynamics is quenched from $T \gg T_c$ to $T_c$, the early times of evolution also display universal behaviour. A new independent exponent, $\theta$, associated with the anomalous dimension of the initial order parameter, was introduced to describe the system behaviour during this {\it critical initial slip} regime. Denoting by $m_0$ the initial magnetization ($0 < m_0 \ll 1$), this regime is found in the time range $t_{mic} < t < m_{0}^{-z/x_0}$, where $t_{mic}$ is some microscopic time and $x_0 = \theta z + \beta /\nu$ ($\beta$ and $\nu$ are the equilibrium critical indices). The magnetization ($m(t) = N^{-1} \sum_{i} <\sigma_{i}>$) increases with time as 
\begin{equation}
m(t) \sim m_0 t^{\theta}.
\end{equation}
$\theta$ is also related to the decay of the auto-correlation function from a disordered initial state,
\begin{equation}
A(t) \sim t^{-\lambda}
\end{equation}
with $\lambda = d/z - \theta$ in $d$ space dimensions. The relation between short time dynamics and damage spreading was recently clarified by Grassberger \cite{grass95}.

For a finite system it is expected \cite{janssen89} \cite{Li95} that the moments of the order parameter, $m^{(k)}$ ($kth$ moment of the magnetization), have the scaling form,
\begin{equation}
m^{(k)}(t, \tau, L, m_0) = b^{-k\beta/\nu} m^{(k)}(b^{-z}t, b^{-1/\nu}\tau, bL, b^{-x_0}m_0)
\end{equation}
where $\tau = (T -T_c)/T_c$ and the initial correlation length is null. 
Setting $\tau=0$ and the arbitrary scaling factor $b \sim t^{1/z}$, one obtains from eq.(3),
\begin{equation}
m(t,L, m_0) \sim t^{-\beta/\nu z} F(t/t_L, t/t_0)
\end{equation}
where $t_L \sim L^{z}$ and $t_0 = m_{0}^{-z/x_0}$.
Following the scaling relations for the magnetization and its higher moments, it is possible to infer that 
the time dependent Binder cumulant \cite{privman},
\begin{equation}
U(t,L) = 1-\frac{m^{(4)}}{3(m^{(2)})^{2}}
\end{equation}
obeys,
\begin{equation}
U(t,L_1) = U(b^{-z}t, L_2)
\end{equation}
for $\tau =0$, $m_0 = 0$ and two system sizes ($L_1$ and $L_2$) with $b=L_{2}/L_1$. The exponent $z$ can be obtained from a data collapse with a time rescaling factor $b^{-z}$. Since only early times are considered, this is a rather efficient method to extract $z$. 
Once $z$ is known, the static exponent $\beta /\nu$ is recovered from a similar scaling analysis of $m^{(2)}$. 

Starting with random initial configurations (with $m_0 = 0$) and following the evolution at $T_c$ of the spin auto-correlation function,
\begin{equation}
A(t) = \frac{1}{N} \left< \sum_{i=1}^{N} \sigma_i (0)  \sigma_i (t)  \right> 
\end{equation}
the power-law decay, eq.(2), is visible after a short transient regime and $\lambda$ (and therefore, $\theta$) can be obtained \cite{okano97}. A direct measurement of $\theta$ is possible from eq.(1) - the samples are then prepared with a sharply defined small value of $m_0$. After a few Monte Carlo steps (MCS) a straight line appears in the log-log plots and $\theta$ is computed from its slope. 

The $2$-state isotropic majority vote model is defined \cite{mario} by a set of 'voters' or 'spin' variables $\{ \sigma_i \}$ taking the values $+1$ or $-1$ and evolving in time by a single spin-flip like dynamics with a probability $W_i$ given by
\begin{equation}
W_i (\sigma) = \frac{1}{2} \left[ 1 - \sigma_i (1-2q) S \left( \sum_{\delta} \sigma_{i+\delta}\right) \right]
\end{equation}
where $S(x) = sign(x)$ if $x\neq 0$, $S(x)=0$ if $x=0$ and the sum is over nearest neighbours of $\sigma_i$. The control parameter $q$ plays the role of temperature in equilibrium systems and measures the probability of aligning antiparallel to the majority of neighbours. In two dimensions this model has a ferromagnetic stationary phase for $0\leq q \leq q_c$ undergoing a second order phase transition to a paramagnetic phase at $q_c$ ($q_c = 0.075(1)$ for a square lattice \cite{mario}, \cite{nos}). The static critical behaviour is Onsager-like \cite{mario}, \cite{nos}, \cite{santos95}. According to Grinstein et al's arguments \cite{GJY}, its dynamic critical behaviour is the same as model A (Ising) \cite{hon} and therefore Janssen et al's renormalization group analysis should also apply to such a nonequilibrium model. The aim of this communication is to report a direct confirmation of the above conjecture from the results of a short time dynamics study of this model. The exponents $z$ and $\theta$ are found to be indistinguishable from the corresponding Ising values, $z=2.172(6)$ \cite{grass95}, $z=2.1665(12)$ \cite{night96} and $\theta = 0.191(3)$ \cite{grass95} and the existance of an intermediate scaling regime is verified.  

Simulations were carried out for square lattices of side $L=16, 32, 64$ and $128$ with periodic boundary conditions. Random initial configurations with $m_0 = 0$ were used in the study of $U(t)$ and $A(t)$, whereas a small excess of plus spins, randomly distributed on the lattice, was taken to produce a selected value of $m_0 \neq 0$. In order to prepare a sample with a precise magnetization and negligeable correlation length, we generated a lattice state with equal probability of occupation for both spin states and then flipped the spin at randomly chosen sites until the desired magnetization was obtained.
The lattice was updated by flipping randomly picked spins \cite{rit97} with probability given by eq.(8) with $q=q_c = 0.075$. The evolution was followed for up to $1000$ MCS. Averages were performed over a large number of histories (up to $4 \times 10^{5}$ independent initial configurations). 

In Figure 1 Binder's cumulant, $U(t,L)$, is displayed against $t/L^{z}$ for different values of system size and initial magnetization $m_0 = 0$. The value of the $z$-exponent obtained from the best collapse was $2.170(5)$, which is in very good agreement with the ones obtained by Grassberger \cite{grass95} from a damage spreading study  and by Nightingale et al \cite{night96} from a new method (variance reducing) Monte Carlo algorithm for the two dimensional Ising model with nonconservative dynamics. 
Figure 2 shows a plot of $log \; m(t)$ against $log \; t$ for $m_0 = 0.03125$ and different system sizes ($L=16$, $32$, $64$ and $128$); for comparison a straight line with slope $0.191$ is also drawn. The dependence of $\theta$ on $m_0$ was analised (see Table I). A linear extrapolation to $m_0 =0$ yields $\theta = 0.191(2)$. It is clear from Figure 2 that $\theta$ can be estimated from the study of very small system sizes. 
\begin{figure}
\epsfxsize=85mm
\epsffile{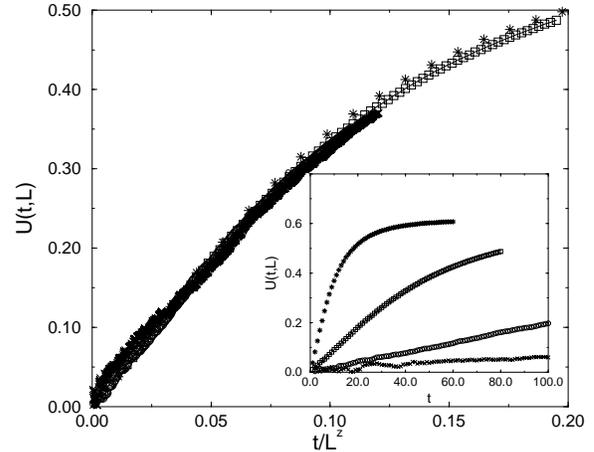}
\caption{The figure shows a collapse plot of the Binder cumulant $U(t,L)$ as a function of $t/L^{z}$ for $L=8 (\ast)$, $16(\Box)$, $32(\circ)$ and $64(\times)$ with $z=2.170$. The inset shows $U(t,L)$ against $t$ (same symbols).}
\end{figure}
\vspace{0.5cm}
In our case, even for $L=16$, the power-law behaviour lasts for two decades of MCS with an exponent close to our best value. This shows that for the measurement of $\theta$ the finite size effects are not important, in contrast with the behaviour of the auto-correlation function, where higher values of $L$ are necessary.  
From Figure 2 we can also obtain an estimate of the crossover time $t_c$ when the magnetization changes to the decreasing power-law behaviour ($t^{-\beta/\nu z}$), before entering the ultimate exponential regime. It is clear that for $t<t_c$ the magnetization presents a power-law increase for all values of $L$ with no significative finite size effects. It is also remarkable that this power-law appears at a very small time ($t_{mic}\sim 1$ MCS). 

In Figure 3 we show a double-log plot of the auto-correlation function, $A(t)$, as a function of time for different values of $L$. To get the critical exponent $\lambda$ we discarded the first 10 MCS. 

\begin{table}
\begin{center}
\caption{
Exponent $\theta$ as a function of $m_0$ for $L=32$. The value obtained for $m_{0} \rightarrow 0$ is $\theta = 0.191(2)$, in good agreement with 
the ones in literature \protect\cite{okano97}, \protect\cite{grass95}.}

\begin{tabular}{c c}
$m_0$          &  $\theta$  \\
\hline
0.023437500    &  $0.1870(20)$  \\
0.031250000    &  $0.1856(20)$  \\
0.058593750    &  $0.1809(20)$   \\
0.080078125    &  $0.1783(20)$ \\
\hline
$m_{0} \rightarrow 0$ & $0.191(2)$
\end{tabular}
\end{center}
\end{table}
\begin{figure}
\epsfxsize=85mm
\epsffile{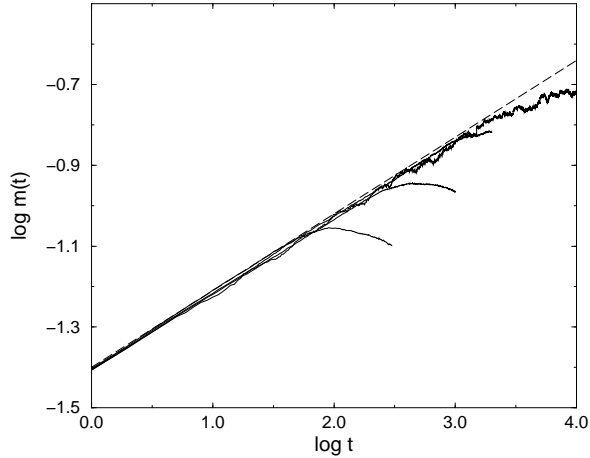}
\caption{Log-log plot of the time evolution of the magnetizatiom $m(t)$ for system sizes $L=16$, $32$, $64$ and $128$ from bottom to top. The initial magnetization was $m_0 = 0.03125$. The dashed guide line has slope $0.191$ and was plotted for comparison.} 
\end{figure}    
The best value obtained for $\lambda$ was $0.735(5)$, which agrees with recently published results \cite{okano97}. The scaling relation $\lambda = d/z - \theta$ is well obeyed for the values obtained before, within the statistical errors.

In summary, we have investigated the dynamic behaviour of a critical nonequilibrium model, the two dimensional majority vote model, making use of the early time dynamic Monte Carlo method. By following the time evolution of the magnetization, Binder cumulant and time auto-correlation function for systems of various sizes, we were able to calculate numerically the values of the exponents $z=2.170(5)$, $\theta = 0.191(2)$ and $\lambda = 0.735(5)$. These values are in very good agreement with their corresponding two dimensional Ising results - a direct confirmation of the stability of the kinetic Ising fixed point with respect to irreversibility of the microscopic rates (of a certain kind). The effect of the absence of detailed balance has probably to be sought in properties like the cluster structure and the dynamics of pattern formation \cite{vinalis}. Another dynamic exponent, $\theta_1$, the global persistent exponent, was recently introduced by Majumdar et al \cite{maj96}. It measures the persistence of the sign of the magnetization and is related to $\theta$ for a Markovian system \cite{maj96}. Evidence of non-markovianicity was reported for a nonequilibrium model \cite{nora} but the situation is unclear in the Ising case \cite{maj96} \cite{sz}. The persistence probability of the majority vote model is currently being investigated. 

{\em Acknowledgments:}
MAS wishes to thank J. Drugowich de Fel\'\i cio for a very helpful discussion. This work was supported by JNICT/PRAXIS XXI (Portugal) under project number: PRAXIS/2/2.1/Fis/299/94. 

\begin{figure}
\epsfxsize=85mm
\epsffile{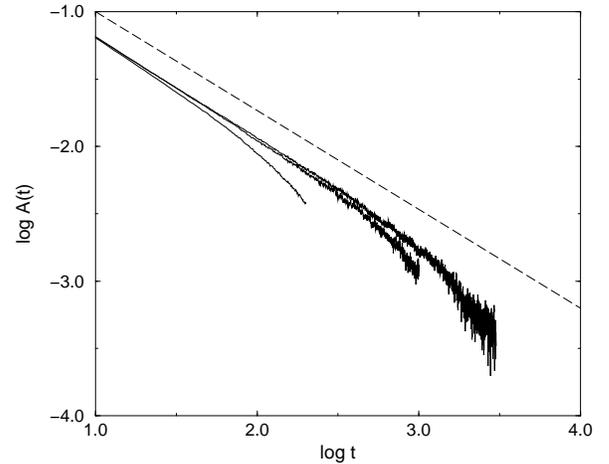}
\caption{Log-log plot of the auto-correlation function $A(t)$ for system sizes $L=32$, $64$ and $128$ (bottom to top) with initial magnetization $m_0 = 0$. The dashed guide line has slope $0.735$.}  
\end{figure}

\end{multicols}
\end{document}